\documentclass[aps,prl,reprint,twocolumn,
superscriptaddress,longbibliography,nofootinbib]{revtex4-2}  
\usepackage{tabularx}
\makeatletter
\def\hlinewd#1{%
\noalign{\ifnum0=`}\fi\hrule \@height #1 %
\futurelet\reserved@a\@xhline}
\makeatother
\usepackage{scrextend}

\usepackage{cellspace} 
\usepackage{amsmath}
\usepackage{amssymb}
\usepackage{epsfig}
\usepackage{graphicx}
\usepackage{hyperref}
\usepackage{dcolumn}   
\usepackage{slashed}
\usepackage{color}
\usepackage{rotating}
\usepackage[table,xcdraw,dvipsnames]{xcolor}
\usepackage[utf8]{inputenc}
\usepackage{colortbl}
\usepackage[normalem]{ulem}
\usepackage{mathrsfs} 

\usepackage[vcentermath]{youngtab}
\usepackage{ytableau}

\usepackage{calligra} 

\definecolor{nicered}{rgb}{0.7,0.1,0.1}
\definecolor{nicegreen}{rgb}{0.1,0.5,0.1}
\definecolor{red}{rgb}{1.0, 0, 0}

\newcommand{\SU}{{\rm SU}}

\newcommand{\bdm}{\begin{displaymath}}
\newcommand{\edm}{\end{displaymath}}
\newcommand{\bea}{\begin{eqnarray}}
\newcommand{\eea}{\end{eqnarray}}

\newcommand{\mB}{\mathcal{B}}
\newcommand{\mN}{\mathcal{N}}
\newcommand{\mM}{\mathcal{M}}

\newcommand{\mC}{\mathcal{C}}

\definecolor{darkgreen}{rgb}{0.1,0.7,0.1}
\definecolor{nicered}{rgb}{0.7,0.1,0.1}
\definecolor{nicegreen}{rgb}{0.1,0.5,0.1}
\definecolor{red}{rgb}{1.0, 0, 0}
\definecolor{niceblue}{rgb}{0,0,0.8}
\hypersetup{colorlinks,citecolor= nicegreen,linkcolor= nicered,urlcolor=nicered}

\def\eq#1{{Eq.~(\ref{#1})}}

\def\gsim{\raise0.3ex\hbox{$\;>$\kern-0.75em\raise-1.1ex\hbox{$\sim\;$}}}
\def\lsim{\raise0.3ex\hbox{$\;<$\kern-0.75em\raise-1.1ex\hbox{$\sim\;$}}}

\def\mb[#1]{\mathbf{#1}}

\definecolor{LightCyan}{rgb}{0.88,1,1}
\definecolor{piggypink}{rgb}{0.99, 0.87, 0.9}
\definecolor{applegreen}{rgb}{0.55, 0.71, 0.0}
\definecolor{darkpastelgreen}{rgb}{0.01, 0.75, 0.24}
\definecolor{green-yellow}{rgb}{0.68, 1.0, 0.18}

\newcommand{\beq}{\begin{equation}}
\newcommand{\eeq}{\end{equation}}
\newcommand{\beqa}{\begin{eqnarray}}
\newcommand{\eeqa}{\end{eqnarray}}

\newcommand{\Sec}[1]{ \medskip \noindent {\sl \bfseries #1}}

\begin{document}



\title{A Casimir bottleneck in primordial large-$\mathcal{N}$ baryon formation}

\author{Luca Di Luzio}
\email{luca.diluzio@pd.infn.it}
\affiliation{\small \it 
INFN Sezione di Padova, Via Francesco Marzolo 8, 35131 Padova, Italy}
\author{Samuele Di Valeriano}
\email{sdivaler@sissa.it}
\affiliation{\small \it SISSA, International School for Advanced Studies, Via Bonomea 265, I-34136 Trieste, Italy}
\author{Enrico Nardi}
\email{enrico.nardi@lnf.infn.it}
\affiliation{\small \it Istituto Nazionale di Fisica Nucleare, Laboratori Nazionali di Frascati, C.P.~13, 00044 Frascati,
Italy}
\affiliation{\small \it Laboratory of High Energy and Computational Physics, NICPB, R\"avala 10, 10143, Tallinn, Estonia}


\begin{abstract}
\noindent
We study baryon $(\mathcal{B})$  formation in the early Universe in a confining $SU(\mathcal{N})$ gauge theory with quarks transforming in the fundamental representation.  Casimir scaling of the confining potential 
implies that, at large $\mathcal{N}$, $\mathcal{B}$ formation is hindered by a  bottleneck: for small quark clusters, representing the initial stages of $\mathcal{B}$ assembly, destruction processes greatly outweigh  formation processes. In  a $\mathcal{B}$-$\bar{\mathcal{B}}$ symmetric plasma,  the relic density of cosmologically stable $\mathcal{B}$'s is set, for large $\mathcal{N}$,  during confinement rather than by annihilation freeze-out.
This affects relic-density estimates for $SU(\mathcal{N})$ dark matter models. 
\end{abstract}

\maketitle

\Sec{Introduction.}  
The interactions among Standard Model (SM) particles are governed by the gauge group $SU(3)_c\times SU(2)_L\times U(1)_Y$. 
If additional gauge groups $SU(\mN)$ with $\mN > 3$ exist in Nature, 
the corresponding states would be confined at a high scale determined by the renormalization-group evolution of the associated
$\beta$-function and, quite naturally, would not appear at  low-energy.
However, it is possible that certain hadronic states of a confining $SU(\mN)$  gauge theory are cosmologically stable, and that the energy density associated with their relic abundance could leave a detectable  imprint as a form of dark matter (DM)~\cite{Antipin:2015xia,Mitridate:2017oky,Appelquist:2015yfa,Kribs:2016cew,Morrison:2020yeg,Asadi:2021yml,Asadi:2021pwo,Gouttenoire:2023roe,Fleming:2024flc,Profumo:2025var}.
Drawing on the QCD analogy, it is natural to assume that $SU(\mN)$  mesons $(\mM)$ possess decay channels, whereas $SU(\mN)$  baryons $(\mB)$ are stabilized by a conserved  baryon number, thereby providing a viable DM candidate. Accordingly, it becomes important to study the formation of $\mB$ baryons in the early Universe. 
In this work, we carry out this study for $SU(\mN)$  theories with large $\mN$.\footnote{
Large $\mN$ gauge models are especially   
motivated in axion physics, as they can protect the Peccei-Quinn symmetry  from explicit breaking by effective operators up to  dimension $\mN$~\cite{DiLuzio:2017tjx,Ardu:2020qmo,Lu:2023ayc,axionquality}.}
Baryon properties in the large $\mN$ limit were first studied 
in Ref.~\cite{Witten:1979kh}
(see  \cite{Manohar:1998xv,Lucini:2012gg} for reviews). One result particularly relevant  for the present work is that the processes of  $\mB$-$\bar \mB$ production or annihilation in mesonic reactions are exponentially suppressed 
as $e^{-c \, \mN}$, with $c$ an $O(1)$ coefficient.
However, in the early Universe, baryons are also formed directly from the quark--gluon plasma during the confinement transition, a process 
to which large-$\mN$ arguments cannot be straightforwardly applied. In QCD,  this results in a baryon-to-meson abundance without any significant suppression~\cite{Braun-Munzinger:2003pwq,Andronic:2011yq,
Andronic:2017pug}.
Whether or not this picture extends to the large-$\mN$ case
is therefore a crucial question.

To model  $\mB$ formation during  the 
$SU(\mN)$ confining phase transition (PT), 
we rely on two key ingredients: 
Casimir scaling~\cite{Ambjorn:1984mb,Deldar:1999vi,Bali:1999hx,Bali:2000un,
Shevchenko:2001ij,Semay:2004br,Cardoso:2011cs,Mykkanen:2012ri}
for the strength of the interaction  potential between color sources, 
and a quark 
recombination mechanism inspired by the QCD Resonance Recombination Model (RRM)~\cite{Ravagli:2007xx,Ravagli:2008rt}. 
We show that, in the large-$\mN$ limit, $\mB$ formation is extremely  suppressed  relative to  meson formation, yielding relic abundances that can be well below those predicted by conventional $\mB$-$\bar\mB$
 annihilation freeze-out scenarios in case  thermal initial conditions are assumed.
The underlying reason is a form of {\em Casimir bottleneck}, analogous to familiar bottlenecks in Big Bang nucleosynthesis, such as the deuterium bottleneck delaying $^4$He formation, and the absence of stable nuclei with $A=5$ and 
$A=8$ that hinders the production of heavier elements.
In the present case, the initial formation of small quark clusters is hindered because destruction processes greatly outweigh cluster formation, whereas meson formation proceeds at full rate, rapidly depleting the pool of free quarks. 
To make the role of $SU(\mN)$  Casimir scaling more transparent,  we consider a simple model with one light quark flavor  $(n_f=1)$,  singlet under the SM gauge group. In this scenario, the mesons are stable and 
can be cosmologically problematic.
However, since the Casimir hierarchy does not depend on the number of quark flavors, the $n_f=1$ benchmark provides a valid  illustrative example. A realistic $n_f=3$ model that incorporates this mechanism, 
provides cosmologically safe  decay channels for the $\mM$-mesons, 
and can lead to relic $\mB$  abundances compatible with the observed DM density is presented in the companion paper~\cite{companion2}.

\Sec{Cornell potential and Casimir scaling.}
In a confining $SU(\mN)$ gauge theory, the interaction between two  static color sources $a$ and $b$ at short and intermediate distances 
is well described by a  Cornell potential~\cite{Eichten:1974af,Eichten:1978tg,Eichten:1979ms}: 
\begin{equation}
    V_{ab}(r) = \hat\mC_{ab}\left(\frac{\alpha}{r} - \sigma\, r \right)\,,
\label{eq:Cornell} 
\end{equation}
where $\alpha$ and $\sigma$ denote, respectively, the Coulomb  and the linear potential coefficients.
According to the  Casimir scaling hypothesis~\cite{Ambjorn:1984mb}, which is exact for the Coulomb term through two loops, receiving the first 
corrections at three-loops~\cite{Anzai:2009tm,Smirnov:2009fh,Anzai:2010td},
and is well supported at intermediate distances 
 by lattice results~\cite{Ambjorn:1984mb,Deldar:1999vi,Bali:1999hx,Bali:2000un,Shevchenko:2001ij,
Semay:2004br,Cardoso:2011cs,Mykkanen:2012ri},
the   color-channel-dependent factors $\hat\mC_{ab}$ are determined by  
 Casimir scaling~\cite{Ambjorn:1984mb}. 
Their normalization is fixed by the reference potential between a quark $\Psi$ and an antiquark 
$\bar\Psi$, for which 
$\mC_{[\Psi\bar\Psi]} = -C_F$, with 
 $C_F$  the quadratic Casimir of the fundamental representation.
At large $\mN$, $C_F \sim \mN$,  so that $\alpha = C_F g_s^2 \sim \mN g_s^2$  is constant  in the 't~Hooft  limit. Therefore,  
   $\hat\mC_{ab}=\mC_{ab}/C_F$, with $\mC_{ab}$ the Casimir factor for the specific color channel,  effectively captures the 
 large $\mN$ dependence of the interaction.
The  coefficients $\mC_{ab}$ can be written in terms of   
the quadratic Casimir    
$C_a$, $C_b$   of the representations of 
the two sources $a$, $b$, 
and of that of the  two-body system $C_{[ab]}$, as~\cite{griffiths_2025_7h9s9-e9h15}: 
\begin{equation}
\label{eq:Cab}
\mC_{ab} = \frac{1}{2}\left(C_{[ab]} - C_a - C_b \right)\,.
\end{equation}
Relevant for the present study are the Casimir of the fundamental,   
$C_\Psi=C_{\bar\Psi}\equiv C_F$, and that 
of a state of $p$-quarks in the color-antisymmetric 
representation: 
\begin{eqnarray}\label{eq:CF}
C_F &=&  \frac{\mN^2-1}{2\mN} \, , \\
\label{eq:Cp}
C_{[\Psi^p]_A} &=& \frac{p(\mN-p)(\mN+1)}{2\mN}\,.
\end{eqnarray}
For a  color-singlet meson  $C_{[\Psi\bar \Psi]}
=0$, 
and hence the  coefficient for binding a quark-antiquark pair into a meson is $\mC_{[\Psi\bar\Psi]}\equiv \mC_{\Psi+\bar\Psi \to [\Psi\bar\Psi]} = - C_F$.
For  $\SU(3)$, the antisymmetric diquark transforms as 
${[qq]_A}\in \bar 3$, and therefore $C_{[qq]_A}=C_F$. 
\eq{eq:Cab} then immediately gives the QCD ``one-half'' rule:
 $\mC_{[q q]_A} =\frac{1}{2} \mC_{[q\bar q]}$.
 
The product $\hat\mC_{ab} \, \alpha$ 
in the Coulomb part of the potential in~\eq{eq:Cornell} acts as an effective coupling governing the interaction between color sources. With the convention adopted in \eq{eq:Cornell},
for positive $\hat\mC_{ab}$ the interaction is repulsive, whereas for  $\hat\mC_{ab}<0$  it is attractive and can lead to bound-state formation. 
For fixed values of $\alpha$ and $\sigma$, 
the corresponding formation cross sections are expected to scale as $\sim |\hat\mC_{ab}|^2$, although for non-static sources  the precise value of the exponent is a kinetic assumption extrapolated from   the perturbative regime. 
Since we are interested 
only in the relative rates of $[\Psi^p]_A$ cluster formation compared to  
$[\Psi\bar\Psi]$ meson formation,  the common dependence on $\alpha$ and $\sigma$  drops out. According to 
\eq{eq:Cornell}, the relevant dependence of the bound-state formation rates is   effectively captured by the Casimir factors. 

We model  $SU(\mN)$ baryon formation by adopting a scenario inspired by the 
QCD RRM~\cite{Ravagli:2007xx,Ravagli:2008rt},  
itself derived from an underlying Boltzmann equation (BE). In the RRM,  meson formation  is implemented as a scattering process in which a quark and an  antiquark recombine into a meson resonance. 
Baryon formation is treated as a two-step process~\cite{He:2019vgs}. 
For $SU(3)$, the first step consists of the recombination of two quarks from the thermal medium into a diquark in the attractive antisymmetric color-antitriplet channel. The  diquark then recombines with a third quark to form a baryon. 
 We assume that for $SU(\mN)$, $\mB$ formation  can  be described  by an analogous stepwise mechanism. 
For large $\mN$, however,   the full network of sequential processes involved  
becomes increasingly complex, since  the number of collision terms grows as  $\sim \mN^{\,2}$.
In this work, we are primarily interested in quantifying the impact of the Casimir bottleneck on the final baryon yield, rather than in tracking the detailed evolution of all intermediate quark-cluster abundances. 
To this end, we adopt a simplified network that nevertheless retains the essential features of the full recombination dynamics.
First, we track only the abundances of  clusters of quarks, since in the absence of a $\Psi$-$\bar\Psi$ asymmetry (see below), the antiquark sector undergoes an identical evolution.
We then keep track only of reactions involving a single free quark or antiquark interacting with clusters of $p$ quarks.
This  accounts only for the transitions 
$[\Psi^p] \to [\Psi^{p\pm1}]$ between 
nearest-neighbors (NN) in cluster space.  
The NN  approximation 
is particularly well justified as long as the number density of free quarks is much larger than that of multiquark clusters.  
Clearly, in doing so we  miss 
the effects of  processes involving two multiquark clusters 
 $[\Psi^p]_A + [\Psi^q]_A  \to   [\Psi^{p+q}]_A\ (p+q\leq \mN)$ 
and $[\Psi^p]_A + [\bar\Psi^q]_A  \to   [\Psi^{p-q}]_A + [\Psi\bar\Psi]^q \ (p\geq q)$ 
(that are instead included  in the network 
studied in the companion paper~\cite{companion2}). 
Nevertheless, as can be seen from Table~\ref{tab:yields}, 
the final baryon yield obtained with this approximation  is remarkably accurate
for $\mN=3$, while 
for $\mN=12,\,15$\footnote{We restrict our study to $\mN=0\,(\mathrm{mod}\,3)$. With the conventional quark hypercharge assignments required to open meson decay channels, only this choice allows electrically neutral baryons~\cite{companion2}.}
it differs  only by  a factor $O(10)$ compared to the beyond NN (BNN) network, in which  all the (maximally attractive) antisymmetric clusters 
channels are included~\cite{companion2}.
Thus, the NN approximation  fully captures  the qualitative suppression mechanism we aim to illustrate.
Note that in our approach, the cluster abundances freeze out once the density of free (anti)quarks falls below a sufficiently low value. This is clearly an artifact of our approximation, which does not treat the long-distance regime dominated by color-string fragmentation effects, that are ultimately responsible for the complete hadronization of the residual 
abundance of 
colored clusters into mesons or baryons.

\begin{figure}[t!]
  \centering
      \setlength{\abovecaptionskip}{4pt}
  \setlength{\belowcaptionskip}{-4pt} 
  \includegraphics[width=0.48\textwidth]{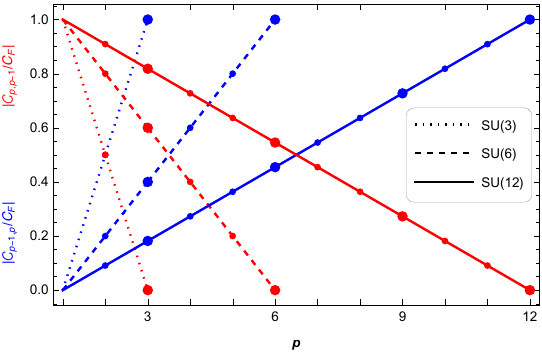}
  \caption{Magnitude of the Casimir factors, normalized to $C_F$, for 
 $p$-quark  cluster formation  
 (blue) and destruction 
 (red), for $\mN=3$ (dotted), $6$ (dashed) and $12$ (solid).}  
  \label{fig:CasimirFactors}
\end{figure}

We  consider the following reactions:
\vspace{-5pt}
\begin{itemize} \itemsep -6pt
\item[-] Direct meson formation:   
\vspace{-2mm}
\begin{equation}
\Psi+ \bar\Psi  \  \to \ [\Psi\bar\Psi]  \,,
\label{eq:mesons}
\end{equation}
\item[-] Destruction of  clusters of $p$-quarks: 
\vspace{-1mm}
\begin{eqnarray} 
\label{eq:PtoPp1}
\Psi+[\Psi^{p}]_A \  &\to&  \ [\Psi^{p+1}]_A \,, \\
\label{eq:PtoPm1}
\bar \Psi+[\Psi^{p}]_A \  &\to&  \ [\Psi^{p-1}]_A + [\Psi\bar \Psi] \,,
\end{eqnarray}
\vspace{-20pt}
\end{itemize}
The process in \eq{eq:PtoPp1} contributes to the formation of $(p+1)$-quark clusters, whereas the process in \eq{eq:PtoPm1} hinders $\mB$ formation by reducing the number of quarks  in the initial cluster, while simultaneously contributing to meson production. 
Using Eqs.~(\ref{eq:Cab})-(\ref{eq:Cp}) 
we can easily derive the Casimir factors 
that weight these reactions:
\begin{align}
\label{eq:CPtoPp1}
&  \mC_{p,p+1} 
 = - \frac{\mN+1}{2\mN} p
\,, \\
\label{eq:CPtoPm1}
& \mC_{p,p-1} 
= 
 - \frac{\mN+1}{2\mN} (\mN-p)\,. 
 \end{align}
 Note that $\mC_{1,0}=  - C_F$
gives the factor that controls direct 
meson formation, while $\mC_{\mN,\,\mN-1}= 0$ 
correctly accounts for the fact that baryons $\mB=[\Psi^\mN]_A$  are $SU(\mN)$ singlets and 
do not feel color forces. 
For consistency we also set 
 $\mC_{\mN,\,\mN+1}  =0$.
In the large $\mN$ limit $|\mC_{1,0}| \sim O(\mN)$, 
while for diquark formation $|\mC_{1,2}|\sim O(\mN^0)$, 
therefore for large $\mN$  the binding 
of quarks into mesons largely dominates. Moreover, for 
the destruction process  that dissociates a diquark producing a meson plus a free quark the color coefficient is $\mC_{2,1}$,  which is also  $O(\mN)$. 
Consequently, the growth of large-$p$ clusters, and ultimately the formation of $SU(\mN)$  baryons, is strongly hindered by a bottleneck that, as can be seen from Fig.~\ref{fig:CasimirFactors}, continues to suppress multiquark cluster formation up to $p<\mN/2$, and is especially severe at small $p$.
Note that the  channel alternative to \eq{eq:PtoPp1} 
in which a mixed-symmetry state 
$[\Psi^{p+1}]_{\rm mix}$ is formed, 
and the  channel alternative to 
\eq{eq:PtoPm1},
in which there is no color contraction 
between $\bar\Psi$ and one $\Psi$,
leading to a mixed state 
$[\bar\Psi\Psi^p]_{\rm mix}$, 
both have positive Casimir coefficient \cite{companion2}, 
$    \mC_{\Psi [\Psi^p]_A \to [\Psi^{p+1}]_\mathrm{mix}} =  \frac{\mN-p}{2\mN}$
and 
$    \mC_{\bar \Psi [\Psi^{p}]_A \to [\bar\Psi\Psi^p]_\mathrm{mix}} =  \frac{p}{2\mN}$,
respectively. 
They correspond to repulsive interactions that cannot yield bound states, and therefore  do not compete with the chain of maximally attractive reactions  in the antisymmetric channels~Eqs.~(\ref{eq:PtoPp1}) and (\ref{eq:PtoPm1}).

\Sec{Baryon formation.}
To estimate the baryon yield,  
 we assume that there is no particle–antiparticle asymmetry in the $\Psi$-sector, and that possible  
 operators capable of transferring the SM asymmetry to  the $\Psi$-sector are absent or sufficiently suppressed to remain out of equilibrium until bound states are formed. 
We  also assume, as a benchmark, 
that hadronization proceeds through a smooth crossover (see \cite{companion2} for  details). 
Accordingly, we study bound-state formation in a homogeneous plasma, without bubble formation or other spatial inhomogeneities.\footnote{Our calculation determines the  $\mB$-$\bar \mB$ symmetric 
component from recombination in a homogeneous PT. If confinement is first order, statistical  fluctuations in the number of $\Psi$ and $\bar\Psi$ in shrinking deconfined pockets add a model-dependent floor~\cite{Asadi:2021yml,Asadi:2021pwo,Gouttenoire:2023roe}. 
This does not change the effects of the Casimir hierarchy, 
that keeps controlling hadronization of the symmetric component,  
but the statistical asymmetry might dominate the total relic abundance.}
The typical time-scale for hadronization is of order 
$\Lambda^{-1}$. 
For a confining temperature  well below the Planck scale, 
$T_c\sim \Lambda \ll m_P$, this  
is much shorter than the characteristic time scales for the evolution of the cosmological scale factor $a$  and of the equilibrium density $n_0$ 
of a particle  with $m \gsim \Lambda$, which are larger by at least a factor $m_P/T_c$:
\begin{equation}
 \frac{\dot a}{a}    = H \sim \frac{T^2_c}{m_{P}}  \,,
\qquad \frac{\dot n_0}{n_0} = -H\left(\frac{m}{T_c}+\frac{3}{2}\right).
\end{equation}
This allows us to neglect the expansion of the Universe during the PT and to keep the equilibrium particle densities fixed at $n(T_c)$. 
In the NN approximation, the equation that controls the evolution of the number density of free quarks $n_{\Psi} = n_{\bar\Psi}\equiv n_1 $ contains a single source term $\bar\Psi + [\Psi^2]_A 
\to [\Psi\bar\Psi] + \Psi$ together with  
a series of destruction interactions 
$\Psi+\bar\Psi \to [\Psi\bar\Psi] $,
$\Psi +[\bar\Psi^2]_A \to [\Psi\bar\Psi]+
\bar\Psi$, etc.~and  
$\Psi +\Psi \to [\Psi^2]_A$,   
$\Psi +[\Psi^2]_A \to [\Psi^3]_A$, etc. 
The assumption of Casimir scaling for the process rates amounts to assuming that each rate scales as $\hat\mC_{p,q}^2$ times a universal factor with dimensions of a cross section, whose precise value is not relevant for the problem at hand and is parametrized, on dimensional grounds, as $1/\Lambda^2$.
Neglecting the cosmic expansion, the  evolution equation for $ n_1 $ can be written as:
\begin{equation}
\label{eq:n1density}
\hspace{-1mm}
    \frac{d n_1}{d\tau} = \frac{n_1}{\Lambda^2} 
    \left[ \hat\mC^2_{2,1} n_2
    - \!\!\sum_{p=1}^{\mN-1} \left(
    \hat\mC^2_{p,p-1} + s_{p}\,\hat\mC^2_{p,p+1}\right) n_{p} 
    \right] \, ,
\end{equation}
where $n_p\equiv n_{[\Psi^p]_A}$. 
The coefficient $s_{p} $  multiplying the second term in parenthesis, which accounts for the formation of a $(p+1)$-cluster, is a spin factor required for consistency. The total wavefunction of the quarks in a cluster must be antisymmetric. Since the color wavefunction is antisymmetric, the combined spin-spatial wavefunction must be symmetric. 
The initial state consists of one free quark, with spin multiplicity  $2 S+1=2$,  and one $p$-cluster. Assuming that $p$-clusters form preferentially in the symmetric
$s$-wave ground state, the $p$ quark spins must be fully aligned, yielding  $S=p/2$. The total spin multiplicity of the initial state is therefore $2(p+1)$. The  final  $(p+1)$-cluster has spin multiplicity $ p+2$, giving a spin factor
\begin{equation}
s_p = \frac{p+2}{2(p+1)} \, , 
\end{equation}
that ranges between $2/3$ for $p=2$ to $\sim 1/2$ for $p \gg 1$. 
Although the $s_p$ factors are not particularly small numbers, the 
cumulative effect in suppressing baryon production is numerically important. 
Note that, for the destruction process involving an antiquark and a $p$-cluster, \eq{eq:PtoPm1}, the spin factor is always unity when the production of both pseudoscalar and vector mesons is taken into account. 
Intuitively, this is because the antiquark can combine with a quark with (anti)aligned spin, producing a (pseudoscalar) vector meson, and leaving $p-1$ quarks in the color-antisymmetric spin-flavor-symmetric configuration (see~\cite{companion2} for details).
\begin{table}[t]
\centering
\setlength{\belowcaptionskip}{-10pt} 
\setlength{\tabcolsep}{7.8pt}
\begin{tabular}{SrScScSc}
\hline
 & $y_\mB^{\rm NN}$ 
 & $y_\mB^{\rm BNN}$ & $y_\mB^{\rm BNN}/y_\mB^{\rm NN}$\\
\hline
$\mN=3$ \ \quad & $ 
5.08\!\times\!10^{-2}$  & $4.48\!\times\!10^{-2}$ &\ $0.88$ \\
$\mN=12$\quad & $ 
4.32\!\times\!10^{-12}$ & $5.34\!\times\!10^{-11}$ & $12.36$\\
$\mN=15$\quad & $ 
1.47\!\times\!10^{-16}$ & $3.07\!\times\!10^{-15}$ & $20.88$\\
\hline 
\end{tabular}
\caption{Baryon pseudo-yields at $u=200$ including only the 
NN  reactions Eqs.~(\ref{eq:PtoPp1})-(\ref{eq:PtoPm1}),
compared to the results obtained by
including all antisymmetric clusters channels~\cite{companion2}. 
}
\label{tab:yields}
\end{table}
Let us now 
 introduce  the dimensionless pseudo-yield for each  type of $p$-cluster   
$y_p = n_p/\Lambda^3$,\footnote{We normalize to $\Lambda^3$ 
rather than to the  entropy density, 
because pseudo-yields  
  directly provide an approximate estimate of
$n_\mB/n_\mM$, given that at temperatures of order $T_c\sim\Lambda$, 
 $n_{\Psi,\bar\Psi} \approx \Lambda^3$, and essentially all quarks and antiquarks 
 recombine into mesons.}
and define the dimensionless 
time variable $u=\Lambda \tau$. 
\eq{eq:n1density} becomes 
\begin{equation}
\label{eq:y1}
\hspace{-2mm}
     \dot y_1 = y_1 
    \left[ \hat\mC^2_{2,1} y_2
    - \!\!\sum_{p=1}^{\mN-1} \left(
    \hat\mC^2_{p,p-1} + s_{p}\,\hat\mC^2_{p,p+1}\right) y_{p} 
    \right] \, ,
\end{equation}
where  $\dot y_1 = \frac{dy_1}{d u}$.    
A generic equation in the baryon synthesis chain, describing the evolution of clusters containing $2\leq p\leq \mN$ quarks, has the following form:
\begin{eqnarray}
    &\dot  y_{p} = y_1 \left[ \frac{1}{1+\delta_{2p}} \,s_{p-1}\,\hat\mC^2_{p-1,p} y_{p-1} 
    +   
\hat\mC^2_{p+1,p} y_{p+1} 
\right.\nonumber  \\ &  \hspace{1.5cm} \left. 
- \left(\hat\mC^2_{p,p-1} + s_p\,\hat\mC^2_{p,p+1}\right) y_{p} \right]\,,
\label{eq:chain}
\end{eqnarray}
where $ \frac{1}{1+\delta_{p2}}$ in the first term accounts for the symmetry factor of $1/2$ that  is required, for $p=2$, to avoid double counting  the $\Psi+\Psi \to [\Psi^2]_A$ reaction.
Note that since baryons $\mB\sim [\Psi^\mN]_A$
are  $SU(\mN)$ singlets and do not participate in color-driven  processes,  $\mC_{\mN,\mN-1} = \mC_{\mN,\mN+1}=y_{\mN+1}=0$.
Consequently, the last two equations of the chain, for  
$\dot y_{\mN-1}$ and $\dot y_\mN$, contain respectively only three and a single collision term.
The structure of Eqs.~(\ref{eq:y1})--(\ref{eq:chain})
may appear  unfamiliar in the context of early Universe 
particle reactions. This is because, within the Casimir weighted recombination approach, inverse reactions are absent. For example, once a meson is formed through the reaction~\eq{eq:PtoPm1}, being a color singlet no longer participates in color-driven recombination/destruction processes. Clearly, this approach neglects several effects, such as the breakup of thermal clusters by gluons, in addition to the string fragmentation effects discussed above. Our numerical estimates should therefore be interpreted with care.

\begin{figure}[t!]
    \centering
    \setlength{\abovecaptionskip}{4pt}
    \setlength{\belowcaptionskip}{-4pt}
    \includegraphics[width=0.48\textwidth]{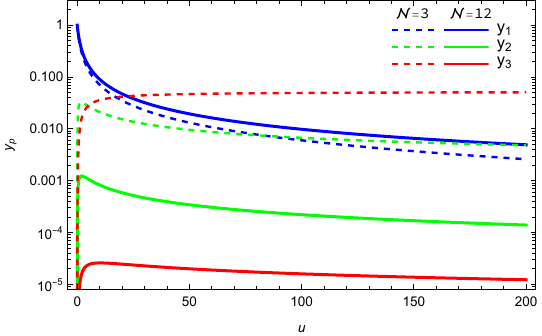}
    \caption{  
     Evolution of the pseudo-yields of free quarks (blue), diquarks (green) and 
     three-quark clusters (red)  for $\mN=3$ (dashed) 
  and $\mN=12$ (solid). 
}
    \label{fig:Evolution}
\end{figure}

Results of integrating the network 
with initial conditions $y_1(0)=1$, $y_p(0)=0$ ($2\leq 
p\leq\mN)$ up to $u=200$ 
are presented in Fig.~\ref{fig:Evolution}.
For symmetric initial conditions $\bar y_1(0)=y_1(0)=1$, 
the same results are obtained for the antibaryon sector.
The evolution of the pseudo-yields for free quarks (blue),  diquarks (green) and three-quark clusters (red) is shown  for $\mN=3$ (dashed lines) 
and $\mN=12$ (solid lines).
For $\mN=3$ the pseudo-yield 
of three-quark clusters (i.e.~ordinary baryons)
grows quickly 
exceeding the abundance of diquarks around $u\sim 5$,
and reaching values of several percent already at $u\sim O(10)$. 
This is in broad agreement with QCD results~\cite{Andronic:2011yq} and confirms that  in $SU(3)$ there is no Casimir bottleneck. 
For $\mN=12$  the situation  changes significantly.
The abundance of free quarks is rapidly depleted due to efficient meson formation, while the abundances of diquark and three-quark clusters are suppressed by approximately four and  five  orders of magnitude, respectively, relative to the initial number density of free quarks, thereby highlighting the significance of the bottleneck.
For $p> 3$ the abundances of all $p$-clusters are increasingly suppressed, resulting in a final $\mB$ pseudo-yield as low as $\sim 10^{-12}$, see Table~\ref{tab:yields}.

\begin{figure}[t!]
  \centering
\hspace{-0.0cm}
  \includegraphics[width=0.48\textwidth]{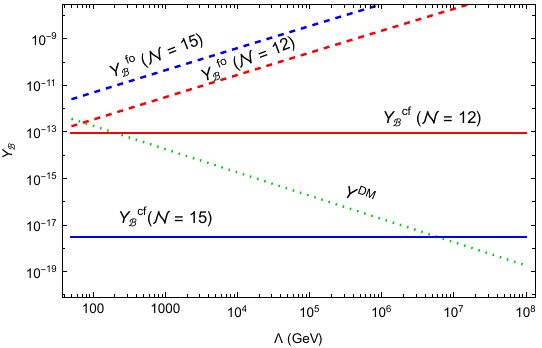}
  \caption{Comparison of the baryon yield from annihilation freeze-out (dashed lines) and from baryon formation at confinement in the NN approximation (solid lines) for 
 $\mN=12$ (red) and $\mN=15$ (blue). 
The   $\mB$-yield  
that reproduces the observed DM density (green dotted line) is included only for reference.}  
  \label{fig:PlotCompare}
\end{figure}

\Sec{Baryon relic density from freeze-out.}
Let us now compare the relic baryon density obtained from the suppression of baryon production due to the Casimir bottleneck,  with that predicted in the more conventional scenario, where   baryons are assumed to be in chemical equilibrium before decoupling, and  the relic abundance is determined by the freeze-out of baryon-antibaryon annihilation. For the latter process, we assume a thermally averaged annihilation cross section
$\langle \sigma v\rangle \sim \frac{\pi}{\Lambda^2} e^{- c\mN}\equiv \sigma_0$, where $ c$ is a coefficient of $O(1)$.  
This expression is motivated by naive dimensional analysis, supplemented by the exponential suppression factor motivated in Ref.~\cite{Witten:1979kh}. The resulting estimates should thus be regarded as only indicative.
Let $Y_\mB=n_\mB/s$  denote the baryon yield  
(normalised to the entropy density $s$), and let us take $x=m_\mB/T$ 
with $m_\mB= \mN\Lambda$ and $g_\mB=\mN+1$ the number of $\mB$ degrees of freedom (dof).  
Assuming radiation domination, constant number of energy  (entropic)  dof 
(including only the SM relativistic dof
$g_*\, (g_{*s}) = 106.75$) and $s$-wave effective annihilation, the conventional equilibrium freeze-out solution is~\cite{Kolb:1990vq}:
\begin{align}
    \hspace{-2mm}
    Y_\mB^{\rm fo} &\simeq \frac{3.8\, x_{\rm fo}}{m_P m_\mB \, \sigma_0  \,g_{*s} g_*^{-1/2} },\\
    \hspace{-2mm}
x_{\rm fo} &\simeq \log\!\left[ 0.04\, m_P m_\mB\,  \sigma_0 \, g_{\mB}g_*^{-{1}/{2}}\right]-\frac{1}{2} \log x_{\rm fo}. 
 \end{align}
To compare $Y_\mB^{\rm fo}$ with the $\mB$ pseudo-yield from  production 
at the confinement epoch, we rescale the pseudo-yield  by the entropy factor: $Y^{\rm cf}_\mB = y_\mB \frac{45}{2 \pi^2 g_{*s}} $.
In Fig.~\ref{fig:PlotCompare},   the 
 rescaled pseudo-yields (solid lines) and the freeze-out yields  (dashed lines) are shown as functions of $\Lambda$, for  $\mN=12$ (red) and $\mN=15$ (blue).\footnote{The results for  $Y_\mB^{\rm fo}$ remain valid as long as  
 $\Lambda < 2.4\times 10^8\,(0.8\times 10^6)\,{\rm GeV}$ for $\mN=12\,(15)$ for which the 
  consistency requirement $x_{\rm fo}> m_\mB/T_c\simeq \mN$  
 remains satisfied.
}
We see that for $\mN=12$, throughout the parameter range   $\Lambda \gsim 20\,$GeV
(corresponding to $m_\mB\gsim 240\,$GeV)
the baryon abundance
produced at the epoch of confinement remains below the value expected from annihilation freeze-out. In other words, baryons are produced in such small numbers that annihilation never becomes efficient.
For  $\mN=15$ the suppression of both baryon production 
and of the annihilation cross section 
are even stronger, so that the same conclusion holds. 
Note that $\mB$ production from mesonic processes in the thermal bath
$\mM\mM\to \mB\bar\mB$, is doubly suppressed:
 by the interaction factor  $e^{-c \mN}$ 
and by an even tinier  Boltzmann factor $e^{-\frac{2m_\mB}{T}} \lesssim 
e^{-2\mN}$. Consequently, its contribution to  $Y_\mB$ is completely negligible.
In  Fig.~\ref{fig:PlotCompare} we also plot 
the value of the baryon yield that would match the observed DM density. 
This corresponds to the dotted green line  and is  obtained from 
\begin{equation}
\hspace{-2mm}
   \frac{1}{2} \Omega_{\scriptscriptstyle \rm DM} h^2 = \frac{m_\mB Y_\mB^{\scriptscriptstyle\rm DM} s_0}{\rho_c/h^2}\; 
    \Rightarrow \;  Y_\mB^{\scriptscriptstyle\rm DM}\approx  2.2 \frac{\mathrm{GeV}}{\mN\,\Lambda} 10^{-10}\!, \!
\end{equation}
where the  factor of $1/2$  counts only the contribution of $\mB$  ($\Omega_{\bar\mB} = \Omega_{\mB}$), and we have used 
$  \Omega_{\scriptscriptstyle \rm DM} h^2 = 0.12$, 
$\rho_c/h^2=8.1\times 10^{-47}\,$GeV$^4$,  $s_0=2.2\times 10^{-38}\,$GeV$^3$ for  the present entropy density and,  in obtaining the final expression, $m_\mB\simeq \mN\,\Lambda$. 
The $\mN=12$ and $\mN=15$ cases then differ only by a factor $5/4$, that is 
negligible on the logarithmic scale shown. We stress that 
the line for $Y^{\rm DM}$ is  only plotted for reference, 
and that its intersections with the  $Y_\mB$ lines 
should not be interpreted as suggesting 
$\mB$ as a DM candidate. This is primarily because, for  $n_f=1$, the contribution of  stable relic mesons 
to $\Omega$ largely dominates over that of baryons. Nevertheless, one qualitative piece of information that can be read off from  Fig.~\ref{fig:PlotCompare} 
is that, 
for large $\mN$,
there exists an interesting range of values of $\Lambda$ for which the relic abundance is determined by the confinement PT,  while remaining compatible with the cosmological limit $\Omega_\mB \leq \Omega_{\rm DM}$.

\Sec{Conclusions.} 
Quark confinement in non-abelian gauge theories remains, so far, an unresolved problem. Consequently, the relative abundance of baryons and mesons emerging from a confining PT in the early Universe cannot be computed from first principles. Nevertheless, some features are well established, such as the Casimir scaling of the interaction strength between unconfined color charges.
Relying on the RRM prescription developed to describe hadronization in QCD, we have exploited this feature to investigate hadron formation in a $SU(\mN)$  theory with large $\mN$. Note that the RRM has been used as a kinetic ansatz, and the validity of its extrapolation from the short $SU(3)$  recombination sequence to an $\mN$-constituent baryon remains an assumption.
We find that baryon formation is strongly suppressed,  so that quarks hadronize almost exclusively into mesons. We have illustrated the underlying mechanism in a simplified 
$n_f=1$ framework, and we have restricted the reactions contributing to baryon and meson formation to the most relevant NN subset. These simplifications are removed in the companion paper~\cite{companion2}, where a more realistic $n_f=3$  model is studied and a BNN reaction network including all the 
most attractive cluster-(anti)cluster channels 
is implemented.  Although the BNN numerical results  can differ by up to one order  of magnitude from the NN approximation, a severe Casimir suppression of baryon synthesis persists. We conclude that this effect is an intrinsic feature of large-$\mN$ $SU(\mN)$ models, and that it should be taken into account when estimating the relic abundances of baryon DM candidates.


\begin{acknowledgments}
\Sec{Acknowledgments.} 
We thank Marco Nardecchia for valuable contributions and constructive discussions throughout the development of this project, and  Giovanni 
Villadoro for useful conversations.  
The work of LDL is supported by the Italian Ministry of University and Research (MUR) via the FIS2 Consolidator Grant project FIS-2023-02106 -- QAXION (CUP: I53C25001880001).
The work of EN is supported  by the Estonian Research Council grant PRG1884
and by the INFN ``Iniziativa Specifica" Theoretical Astroparticle Physics (TAsP).
Partial support from  the Estonian Research Council grants TARISTU24-TK10, TARISTU24-TK3,  CoE grant TK202 “Foundations of the Universe”,
and from the CERN and ESA Science Consortium of Estonia, grants RVTT3 and RVTT7 is also acknowledged.
\end{acknowledgments}

\bibliography{bibliography}

 \end{document}